\documentclass{llncs}
\usepackage[pdftex]{graphicx}
\usepackage[numbers,sectionbib]{natbib}

\begin{document}

\title{Applying Compression to\\a Game's Network Protocol}
\author{Mikael Hirki}
\institute{Aalto University, Finland\\
\email{mikael.hirki@aalto.fi}}
\maketitle

\begin{abstract}
This report presents the results of applying different compression algorithms to the network protocol of an online game. The algorithm implementations compared are zlib, liblzma and my own implementation based on LZ77 and a variation of adaptive Huffman coding. The comparison data was collected from the game TomeNET. The results show that adaptive coding is especially useful for compressing large amounts of very small packets.
\end{abstract}

\section{Introduction} \label{sec:intro}

The purpose of this project report is to present and discuss the results of applying compression to the network protocol of a multi-player online game. This poses new challenges for the compression algorithms and I have also developed my own algorithm that tries to tackle some of these.

Networked games typically follow a client-server model where multiple clients connect to a single server. The server is constantly streaming data that contains game updates to the clients as long as they are connected. The client will also send data to the about the player's actions.

This report focuses on compressing the data stream from server to client. The amount of data sent from client to server is likely going to be very small and therefore it would not be of any interest to compress it.

Many potential benefits may be achieved by compressing the game data stream. A server with limited bandwidth could theoretically serve more clients. Compression will lower the bandwidth requirements for each individual client. Compression will also likely reduce the number of packets required to transmit larger data bursts.

Compressing these data streams poses some requirements on the compression algorithm. The algorithm has to be able to compress data packets of arbitrary length so that they can also be decompressed individually. Especially large numbers of small packets are likely to pose a challenge for the compression algorithms.

Two publicly available and widely-used compression algorithm implementations and my own compression algorithm implementation are compared in this report. Zlib~\cite{zlib} is the traditional compression library which is also widely used for network stream compression. Liblzma~\cite{xz-utils} implements the LZMA algorithm which in general achieves 30\% better compression than zlib. However, there are very few instances where the LZMA algorithm has been applied to stream compression.

The sample data used in this report has been collected from a game called TomeNET~\cite{tomenet}. TomeNET is a multi-player rogue-like game. The graphics are based on drawing characters on the screen. The network protocol used by the game is described in Section~\ref{sec:network-protocol}.

\section{The Network Protocol} \label{sec:network-protocol}

TomeNET uses its own custom network protocol to communicate between the server and the client. The server will write game updates to a buffer and this buffer is flushed at the end of each internal turn. Typically the server runs at 60 turns per second but players can typically perform only a few actions per second at most. The game runs in real-time so events do occur without player involvement in the game world and these may cause updates to be sent to clients.

The protocol is highly structured. One byte always specifies the type of game packet that follows. Most packets consist of a series of binary-encoded numbers that can be 1, 2 or 4 bytes long. The game packets can also contain textual strings.

The protocol consists of over a hundred different game packet types. These packets can instruct the client to draw one or more characters on the screen, request action from the player or transmit various player stats. Typically most of the packets are related to the movement of the player.

The protocol contains multiple sources of redundancy. One of these has actually been eliminated in the design of the protocol. The server can transmit entire rows of characters to be drawn on the screen in the client. These rows are compressed using a form of run-length encoding (RLE).

Other sources of redundancies are the messages generated by the game. For instance, a message is generated every time the player hits a monster or the player is hit by a monster. The player's stats are also retransmitted when at least one of them has changed. Thus the packets may contain unchanged data.

\section{Compression Methods} \label{sec:compression-methods}

Three different compression algorithms are used to compress the data: DEFLATE~\cite{DEFLATE}, LZMA~\cite{LZMA} and an algorithm devised by me. The DEFLATE implementation used is zlib~\cite{zlib} and the LZMA implementation used is liblzma (a part of the XZ Utils package~\cite{xz-utils}).

The DEFLATE algorithm is a combination of the LZ77~\cite{Ziv77auniversal} compression algorithm and Huffman coding~\cite{Huffman52}. The LZ77 algorithm eliminates redundancy in the input data by removing repeated strings. The Huffman coding algorithm creates an optimal prefix encoding based on the occurrence frequencies of the symbols in the input data.

The DEFLATE algorithm supports dynamic Huffman codes. These dynamic codes are used to encode singular blocks. The code for each block is defined in the header by specifying the code lengths for the symbols in the block header. The code lengths themselves are also compressed using a Huffman code.

The LZ77 algorithm creates a parse of the input string. The parse consists of tuples of the form (distance, length, literal). Each tuple describes a possible repeated occurrence of some substring followed by some literal in the input string. Distance specifies the location of the previous occurrence of the substring and length is the length of the occurrence.

The LZMA algorithm is also loosely based on LZ77. LZMA supports huge dictionary sizes and multiple dictionary search structures. The output is encoded with a very sophisticated range encoder.

The algorithm devised by me is based on LZ77. The LZ77 parse is encoded using a variation of adaptive Huffman coding. Normally adaptive Huffman codes~\cite{Vitter:1985:DAD:1382438.1382861} (also sometimes called dynamic Huffman codes) update the coding tree after each symbol has been encoded. In my variation, the coding tree is only updated when a special symbol is transmitted.

The implementation uses suffix trees to compute the LZ77 parse. Ukkonen's algorithm~\cite{Ukkonen95on-lineconstruction} is used to construct suffix trees in linear time. Suffix trees are constructed for blocks of input data. Only a certain number of suffix trees for the most recent blocks are kept in memory. These most recent blocks correspond to the sliding window in LZ77. The parse tuples have the form (position, length, literal) where position is the distance measured in bytes from the beginning of earliest block of data still being kept in memory.

The LZ77 parse is encoded using three Huffman coding tables. They are used to encode the positions, lengths and literals of the tuples. The table used to encode positions also contains special symbols for table rebuilding and block termination. Block in this case is simply a sequence of encoded tuples, followed by the block termination code and padded with zero bits so that it ends on a byte boundary.

All symbols in the tables have a very low weight by default. The tables are initially constructed with these default weights. The algorithm maintains a list of fixed number of earlier tuples. The tables are rebuilt periodically based on this list. A special position code is encoded to the output before the tables are rebuilt.

The coding tables are rebuilt for the first time after two tuples have been emitted. The rebuild interval is doubled until it reaches a fixed value. This allows the algorithm to quickly adapt to the initial data.

\begin{figure}[t]
\begin{center}
\includegraphics[width=10cm]{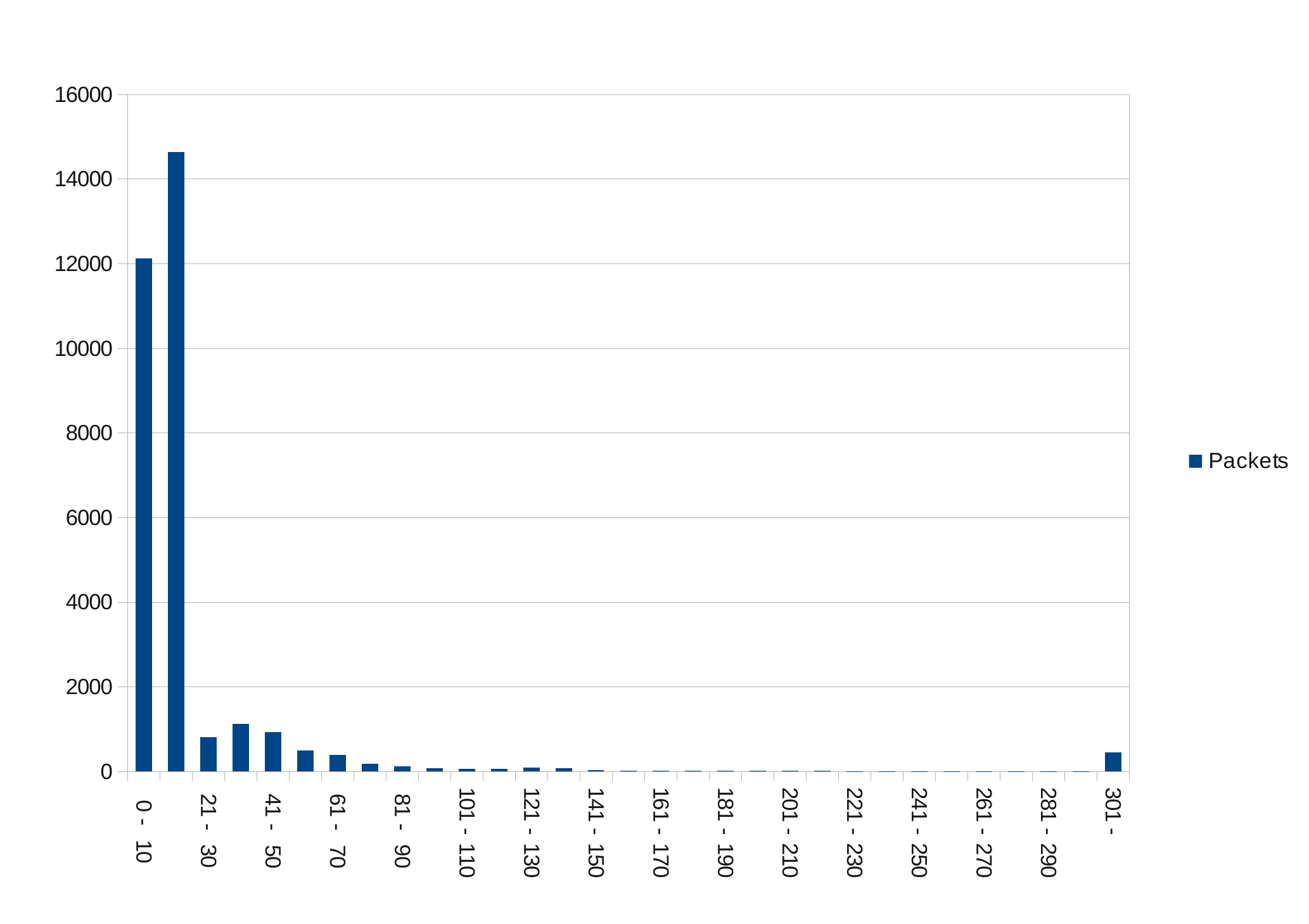}
\end{center}
\caption{The distribution of packet lengths in the sample data. Each vertical bar represents a specific range of packet lengths. The two left-most bars represent the packets whose lengths are in the ranges 0--10 and 11--20}
\label{fig:distribution}
\end{figure}

\section{Compression Results} \label{sec:compression-results}

The sample data consists of data sent by the server during circa 30 minutes of actively playing the game. During this time the server sent nearly 32,000 packets adding up one megabyte of data. Most of this data is binary. Only about 11\% of it is text. The data packets were captured just before they were sent to the operating system's socket layer.

The compression algorithms were forced to compress each data packet so that it could have in theory been sent over the network and decompressed at the client. This was achieved by using the \texttt{SYNC\_FLUSH} option with zlib and liblzma.

The parameters for the algorithms were tuned to obtain the best possible compression result for this data. The compression level was set 9 for zlib and to 3 (with the extreme option disabled) for liblzma.

Zlib achieved an overall average compression ratio of 45\%, liblzma achieved a compression ratio of 48\%, and my compression algorithm achieved a compression ratio of 56\%.

The compression ratio for liblzma actually becomes progressively worse if compression level is increased beyond 3. At compression level 9, the compression ratio is 57\%. If the extreme compression option is enabled, the ratio becomes 65\%. The compression level influences the dictionary size and other internal parameters such as the choice of search structure for the dictionary.

\begin{figure}[t]
\begin{center}
\includegraphics[width=10cm]{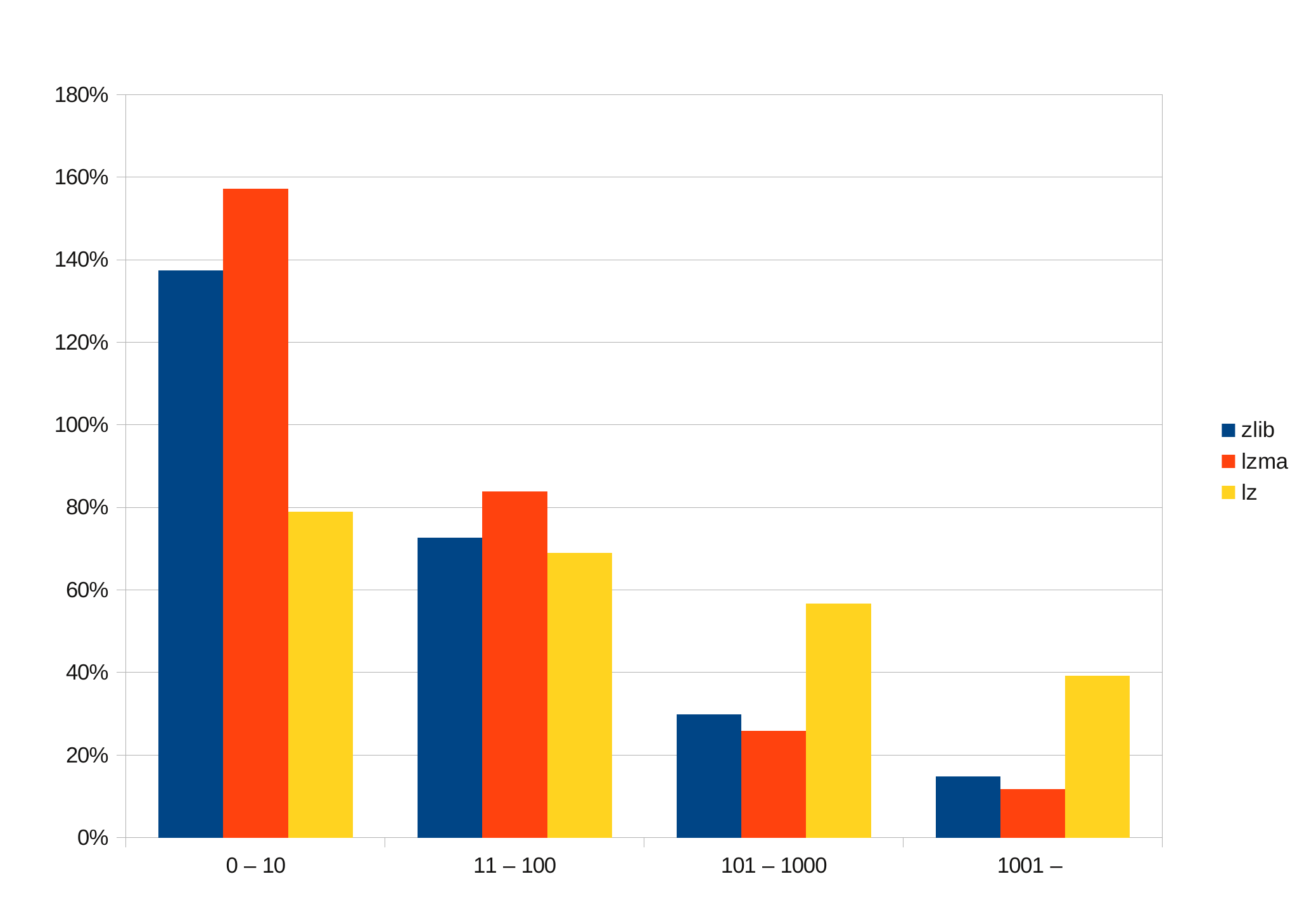}
\end{center}
\caption{The compression ratios of the algorithms for different packet lengths. The packets were divided into four categories based on their lengths: very small (0--10 bytes), small (11--100 bytes), medium (101--1000 bytes), and large (over 1000 bytes). My algorithm is called lz here. Smaller compression ratio is better.}
\label{fig:ratios}
\end{figure}

A vast majority of the packets are extremely small which can be seen in Figure~\ref{fig:distribution}. 84\% of the packets are smaller than 21 bytes and 38\% of them are smaller than 11 bytes. Figure~\ref{fig:ratios} shows how well each algorithm compresses packets of specific lengths. The data presented in this figure was obtained by dividing the packets into four categories based on their lengths. The average compression ratios were calculated based on how much output the packets belonging to each class generated during the entire compression process.

\section{Analysis} \label{sec:analysis}

Zlib achieved the best overall compression ratio in this comparison. However, the difference between zlib and liblzma was only 3 percentage points. This difference is mostly likely due to the fact that liblzma adds more overhead to the packets. This is also evident in Figure~\ref{fig:ratios} where liblzma inflated the smallest packets more than zlib.

Liblzma is a viable alternative to zlib when doing stream compression. Especially if the packets are larger, liblzma will likely achieve a better compression ratio than zlib. The results presented in this report also support this. However, the compression level and other parameters may need to be tuned to suit specific applications.

Liblzma does have one disadvantage which is its memory usage. Although not measured in this report, the memory usage of liblzma is likely far greater than that of zlib. This is a major disadvantage in a server-client setting where there can be hundreds of clients.

Figure~\ref{fig:ratios} also shows that my compression method was the only one to successfully compress the extremely small packets. Both zlib and liblzma actually inflated these packets instead of compressing them. In the case of zlib, this is most likely due to the overhead added by the dynamic Huffman codes. This shows that adaptive coding techniques are especially useful for compressing extremely small packets.

The adaptive Huffman coding used here is actually not practical. Even though the tables are rebuilt on at fixed intervals, the time required to rebuild the Huffman coding tables is still excessive. A more practical alternative would be adaptive arithmetic coding~\cite{Ryabko_Fionov_1999}.

Another problem with the adaptive coding method is that the coding tables are rebuilt at fixed intervals which prevents fast reaction to changes in the input data. This issue could also be solved by using adaptive arithmetic coding.

The use of suffix trees in LZ77-based compression has been studied before by Fiala and Green~\cite{Fiala:1989:DCF:63334.63341}. They show how the suffix tree can be modified to maintain a sliding window. Senft~\cite{Senft:2006:CST:1126009.1126040} noticed that the longest substring matches can be obtained as a by-product of the suffix tree construction algorithm. He has developed several compression methods based on the idea of describing the construction of the suffix tree.

\section*{Acknowledgments}

Thanks to Travis Gagie for his inspirational lectures.

\raggedright
\bibliographystyle{plainnat}
\bibliography{report}

\end{document}